# Temperature Dependent Interaction Non-Additivity in the Inorganic Ionic Clusters

Vitaly V. Chaban[1,2] and Oleg V. Prezhdo[2]

[1] Instituto de Ciência e Tecnologia, Universidade Federal de São Paulo, 12231-280, São José dos Campos, SP, Brazil

[2] Department of Chemistry, University of Southern California, Los Angeles, CA 90089, United States

**Abstract**. Interaction non-additivity in the chemical context means that binding of certain atom to a reference atom cannot be fully predicted from the interactions of these two atoms with other atoms. This constitutes one of key phenomena determining an identity of our world, which would have been much poorer otherwise. Ionic systems provide a good example of the interaction non-additivity in most cases due to electron transfer and delocalization effects. We report Born-Oppenheimer molecular dynamics (BOMD) simulations of LiCl, NaCl, and KCl at 300, 1500, and 2000 K. We show that our observations originate from interplay of thermal motion during BOMD and cation nature. In the case of alkali cations, ionic nature plays a more significant role than temperature. Our results bring fundamental understanding of electronic effects in the condensed phase of ionic systems and foster progress in physical chemistry and engineering.

**Introduction**

The term *ion* was coined by Faraday in 1834 to name species, which move from one electrode to another in the aqueous medium. Faraday did not know chemistry of ions only possessing information about his initial reactants. Ions are omnipresent in the inorganic nature, biology, and lots of human technologies.[1-6] Many metals readily get rid of one, two or even three of their valence electrons. In turn, monoatomic anions are normally singly charged. Doubly charged anions are achieved when a few non-metallic atoms are combined, such as $SO_4^{2-}$, $CO_3^{2-}$, $C_2O_4^{2-}$. However, since ions are highly reactive (chemically and physically), they rarely occur in the free form, except in flames, lightning, electrical sparks and other plasmas. The gas-like ions rapidly bind with counter-ions of suitable charge to generate neutral molecules and crystalline salts. Ions can also be produced in the condensed phases thanks to polar solvents, which weaken the Coulombic cation-anion attraction. Solvated individual ions can be more stable than the corresponding ion pairs as a result of enthalpic and entropic factors. These stabilized species are more commonly found in the environment at low temperatures. Ionic liquids and their mixtures with molecular liquids constitute a vibrant and currently very active research field.[1,7-10]

Ions in crystal lattices of purely ionic compounds are spherical. However, if the positive ion is sufficiently small and highly charged, it distorts the electron cloud of the negative ion. The polarization of the negative ion leads to an extra charge density between the two nuclei. Shared charge density constitutes a contribution to chemical bond, which is known as partial covalence. The larger is an anion, the more polarizable it is. The described effect is present in all ionic systems. It becomes critical when cations possessing high charges (for instance, $Al^{3+}$) are involved. Lithium and beryllium cations are able to compete with aluminium thanks to their tiny radii. For instance, lithium iodide exhibits many common features with covalent compounds. Electronic polarization (interaction non-additivity) must be clearly distinguished from an ionic polarization effect, which is the displacement of ions (nuclei) in the lattice following application of external electric field.

Non-additive interactions, such as electronic polarization and charge transfer, influence heavily physical properties of many molecular liquids, ionic melts, and ionic liquids.[11-13] Although non-additive interactions in the condensed state have been attended before, their dependence on temperature has never received comprehensive attention. The phenomenon has a crucial impact on the phase diagram and numerous phase dependent physical chemical properties. The present work analyzes electronic properties of simple alkali chlorides as a function of temperature. We found that temperature has little impact on electron delocalization, while it significantly influences dipole moments. The dipoles grow because of thermal expansion and geometry fluctuations. Our ab initio results provide important information to develop and test empirical force fields for solid and molten states of ionic compounds.[14]

**Simulation Methodology**

The electronic energy levels and their populations were obtained for inorganic ionic clusters — LiCl, NaCl, KCl containing 20 ions each (Figure 1) — by means of density functional theory (DFT). The wave functions were constructed using BLYP,[15,16] which is a well-established, reliable exchange-correlation functional in the generalized gradient approximation. The wave functions were expanded using the LANL2DZ basis set with effective potentials for core electrons.[17,18] This basis set and this DFT method in combination with one another provide a reasonable tradeoff between accuracy and computational expense. The wave function convergence criterion was set to $10^{-6}$ Hartree for all calculations. Table 1 presents basic properties of all considered systems.

More accurate calculations would require a larger basis set and a hybrid DFT functional. Pure DFT functionals, such as BLYP,[15,16] tend to overestimate electronic polarization by favoring delocalized electrons. The use of the moderate size basis set counteracts this tendency. Further, our calculations show that temperature has a minor impact on electron delocalization. By overestimating electron delocalization, pure DFT functionals should also overestimate temperature-induced changes in delocalization. Hence, more rigorous and computationally intense calculations with hybrid functionals should confirm our conclusion.

Table 1. Simulated systems, basis properties, and selected results. The listed dipole moments and point charges were averaged throughout the corresponding BOMD trajectories at finite temperature

| System | # explicit electrons | # basis functions | Reference temperature, K | Dipole moment, D | Hirshfeld charge/ion, e | ESP charge/ion, e |
|---|---|---|---|---|---|---|
| [Li][Cl] | 100 | 680 | 300 | 3.55±0.02 | 0.34 | 0.75 |
| [Li][Cl] | 100 | 680 | 1500 | 9.66±0.06 | 0.38 | 0.76 |
| [Li][Cl] | 100 | 680 | 2000 | 13.1±0.09 | 0.40 | 0.76 |
| [Na][Cl] | 80 | 560 | 300 | 6.92±0.04 | 0.39 | 0.85 |
| [Na][Cl] | 80 | 560 | 1500 | 11.5±0.07 | 0.45 | 0.83 |
| [Na][Cl] | 80 | 560 | 2000 | 19.7±0.09 | 0.47 | 0.85 |
| [K][Cl] | 160 | 740 | 300 | 11.1±0.04 | 0.48 | 0.87 |
| [K][Cl] | 160 | 740 | 1500 | 20.4±0.13 | 0.54 | 0.85 |
| [K][Cl] | 160 | 740 | 2000 | 23.9±0.09 | 0.55 | 0.85 |

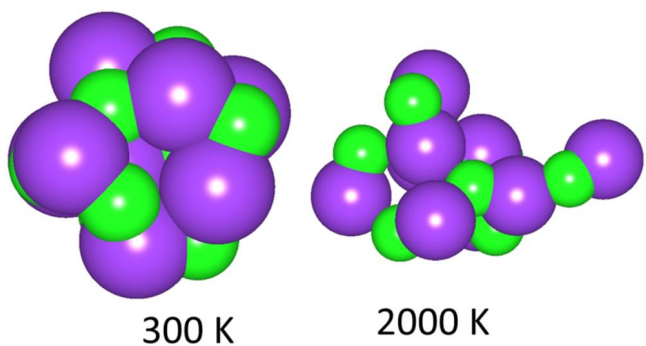

Figure 1. Equilibrium configurations of ionic clusters at room (300 K) and elevated (2000 K) temperatures. The ionic geometry at 2000 K looks significantly perturbed due to thermal motion. Compare dipole moments, which are provided in Table 1. Geometries of KCl are provided as an example. Geometries of LiCl and NaCl are completely analogous.

Prior to the finite-temperature molecular dynamics simulations, the geometries of ionic clusters were optimized using the conjugate gradient geometry optimization algorithm. Therefore, we avoid unnatural forces during the first iterations, which may violate total energy conservation in the simulated systems. The nuclear trajectories were propagated according to the Born-Oppenheimer approximation, whereby energy exchange between electrons and nuclei is neglected. The distribution of electronic density was recalculated at every time-step, whereas atomic coordinates were updated following the Verlet integration scheme for the Newtonian (classical) equations-of-motion. The integration time-step was set with respect to external temperature constituting 2.0, 0.5, and 0.1 fs at 300, 1500, and 2000 K, respectively. The corresponding temperature was maintained through direct velocity rescaling every 50 time-steps ion all BOMD simulations.

The dipole moments (Figure 2) were directly calculated from electronic wave functions. The atomic charges were generated along the MD trajectories using the Hirshfeld scheme[19] (Figure 3) and via electrostatic potential (ESP) fitting (Figure 4) at every BOMD time-step. The atom spheres were defined according to the Merz-Kollman scheme[20,21] utilizing atomic radii from the universal force field, UFF, by Goddard and coworkers.[22] The distance pair correlation functions (PCFs, Figures 5-6) characterize structures of ionic clusters at 300 and 1500 K. The ionic configurations at the last (5000th) BOMD step were optimized to remove a thermal motion effect and the distribution of ESP charges was generated (Figure 7).

The electronic structure computations were performed using GAUSSIAN'09, revision D (*www.gaussian.com*). The subsequent analysis was performed using simple program codes developed by V.V.C.

**Results and Discussion**

Evolutions of dipole moments of the LiCl, NaCl, and KCl ionic clusters at 300, 1500, and 2000 K are provided in Figure 2. The canonically averaged dipole moments and their standard deviations are summarized in Table 1. Note that these dipole moments correspond to entire clusters, instead of a single ion pair. This fact must be kept in mind when comparing our results with other experimental and theoretical investigations. Temperature increase results in a drastic increase of dipole moment. This occurs largely due to thermally induced geometry disturbances rather than due to electrostatic potential alterations, as it will be demonstrated in the following discussion. Increase of cation mass systematically increases dipole moment. The scale of this effect is generally comparable with the thermally induced increase. According to conventional chemical wisdom, the degree of covalence of the ionic bond decreases in the row $Li^+ > Na^+ > K^+$, thus increasing ionicity. Ionicity, in turn, implies larger fractional charges on the cation and on the anion.

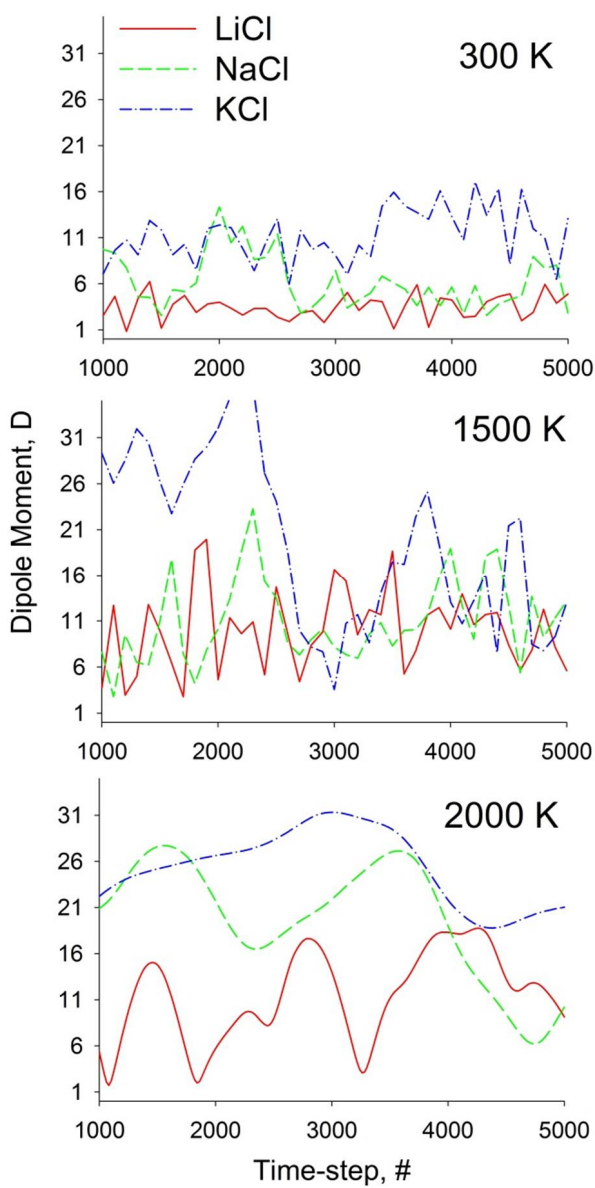

Figure 2. Evolution of dipole moments of the considered ionic clusters during BOMD simulations at 300, 1500, and 2000 K: LiCl (red solid line), NaCl (green dashed line), and KCl (blue dash-dotted line).

Hirshfeld charges (Figure 3) constitute a straightforward and important measure of electron density localization within certain distance around atomic nucleus. Consideration of Hirshfeld charges at the neighboring atoms allows estimating non-polar, polar, ionic and other types of chemical bonding between them. The electron density can be compared for molecular fragments separately and together. If the valence electrons are not shared, then no chemical bonding exists. Evolution of Hirshfeld charges at all temperatures indicates that all systems are in thermodynamic equilibrium. The charges depend both on the cation nature and on the cluster temperature. In concordance with dipole moments, higher temperature brings higher fractional charge. A significant admixture of covalent bonding was detected in the $Li_{10}Cl_{10}$ ionic cluster.

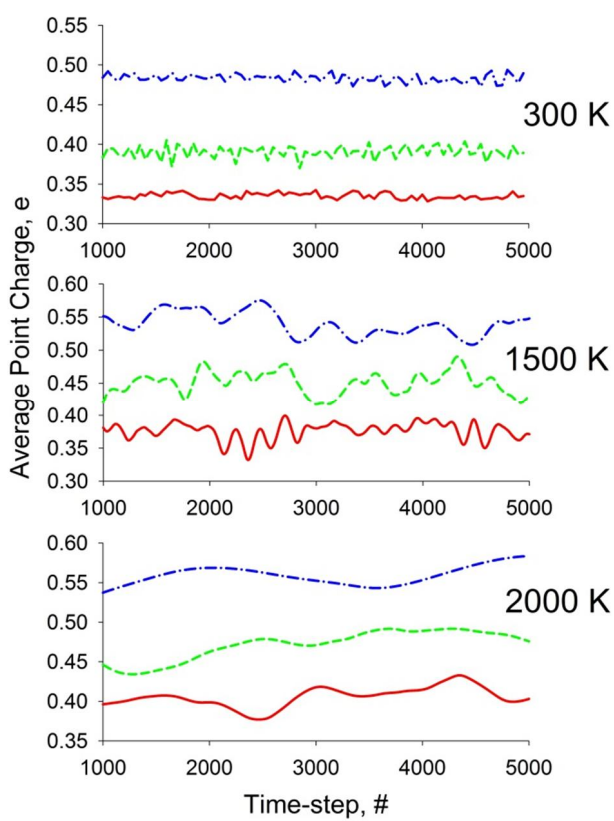

Figure 3. Evolution of Hirshfeld charges (normalized per cation) of the considered ionic clusters during BOMD simulations at 300, 1500, and 2000 K: LiCl (red solid line), NaCl (green dashed line), and KCl (blue dash-dotted line).

As opposed to Hirshfeld charges, ESP charges (Figure 4) reflect electrostatic potential at the surface of the ionic cluster. The corresponding charges, in the same units, are significantly different from Hirshfeld charges exhibiting stronger fluctuations as a result of thermal motion. The charges in $Na_{10}Cl_{10}$ and $K_{10}Cl_{10}$ are quite similar at all temperatures suggesting comparable behavior of the corresponding ions in real systems. The ESP charges in $Li_{10}Cl_{10}$ appear systematically smaller due to contribution of lithium and chlorine into covalent bond. Upon temperature increase, the Li-Cl bonds become more ionic.

Binding energies computed for optimized geometries of ionic clusters confirm our hypothesis about covalent bonding. Compare, the potential energy keeping $Li_{10}Cl_{10}$ as a whole amounts to 7771 kJ mol$^{-1}$, which significantly exceeds values for $Na_{10}Cl_{10}$ (6696 kJ mol$^{-1}$) and $K_{10}Cl_{10}$ (5860 kJ mol$^{-1}$). The basis set superposition errors amount to 1112 kJ mol$^{-1}$ ($Li_{10}Cl_{10}$), 1100 kJ mol$^{-1}$, ($Na_{10}Cl_{10}$), and 1091 kJ mol$^{-1}$ ($K_{10}Cl_{10}$) thus being very similar to one another. Note that all reported binding energies correspond to the respective simulated systems. That is, the energies are given per mole of ionic clusters, not per mole of individual monoatomic ions.

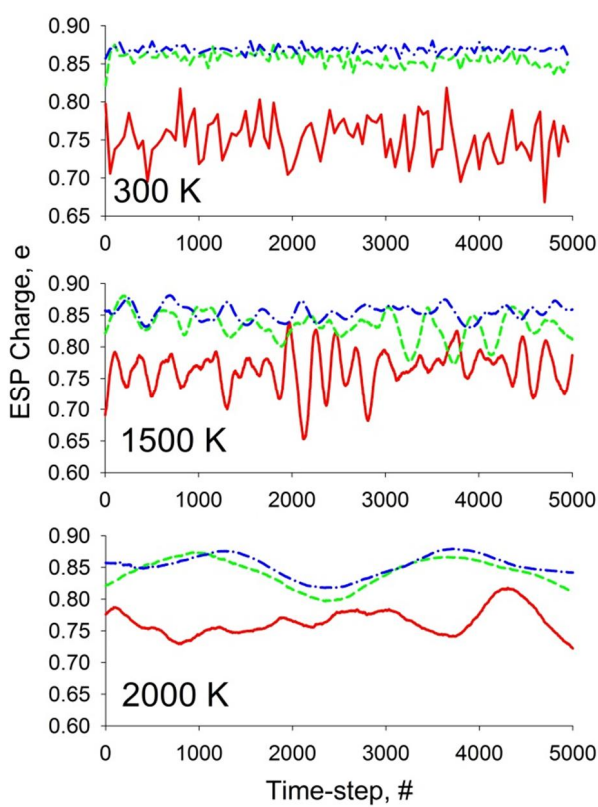

Figure 4. Evolution of electrostatic potential fitted point charges (normalized per cation) of the considered ionic clusters during BOMD simulations at 300, 1500, and 2000 K: LiCl (red solid line), NaCl (green dashed line), and KCl (blue dash-dotted line).

Pair correlation functions were computed for cation-anion, cation-cation, and anion-anion atom pairs in the $Li_{10}Cl_{10}$, $Na_{10}Cl_{10}$, and $K_{10}Cl_{10}$ ionic clusters at 300 K (Figure 5) and 1500 K (Figure 6) using the generated BOMD trajectories. These PCFs provide important information about internal ionic structures in these clusters, which are responsible for electronic and other physical properties. Note that the number of ions in each cluster, 20, is well below thermodynamic limit (several hundred ions). Therefore, assignment of phases, in analogy to the macroscale systems, should not be performed. Since the simulated systems (clusters) are not periodic, surface effects play a drastic role in their structures. The quantum confinement effect is observed when the particle size is small as compared to the electron wavelength. The small overall size confines motion of randomly moving electron to restrict its motion within specific energy levels. The corresponding band gaps widen up and the valence orbital energies also somewhat increase. This fosters chemical reactivity of such entities.

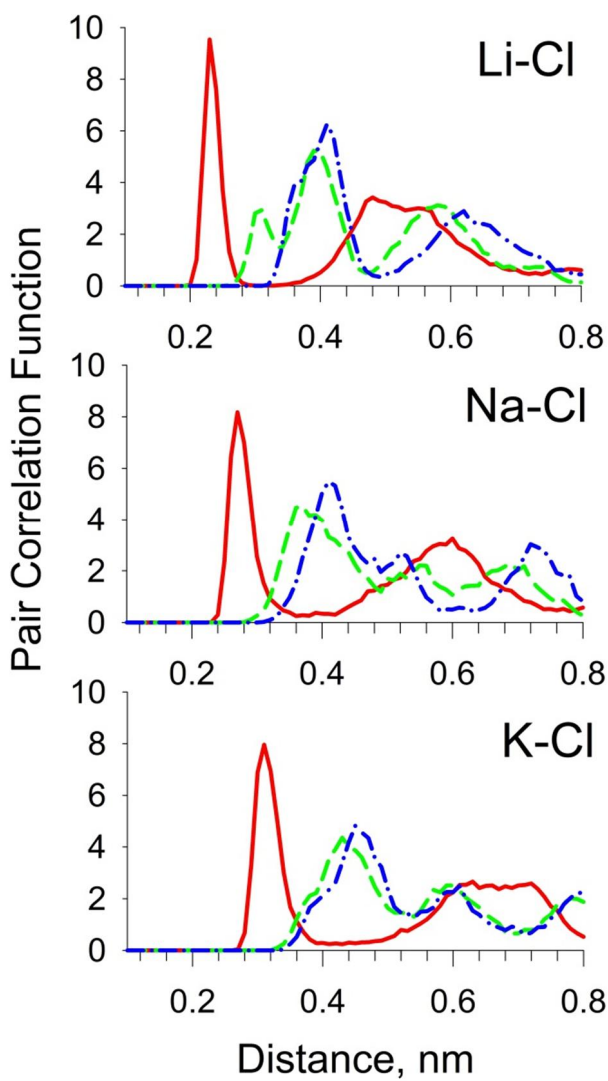

Figure 5. Pair correlation functions in the simulated ionic clusters derived at 300 K: cation-anion (red solid line), cation-cation (green dashed line), and anion-anion (blue dash-dotted line).

Not only cation-anion PCFs are well-shaped driven by Coulombic attraction, but also the first peaks for the cation-cation and anion-anion distance correlations can be distinguished. The anion-anion distances are somewhat larger than cation-cation ones at 300 K, whereas they equalize at 1500 K as a result of thermal motion. The second maxima are smashed, which is a consequence of small cluster sizes. Examination of surface effects extends beyond our present schedule. The depicted PCFs generally support all of our above conclusions including a strong covalent admixture in LiCl and principal similarity of NaCl and KCl irrespective of temperature.

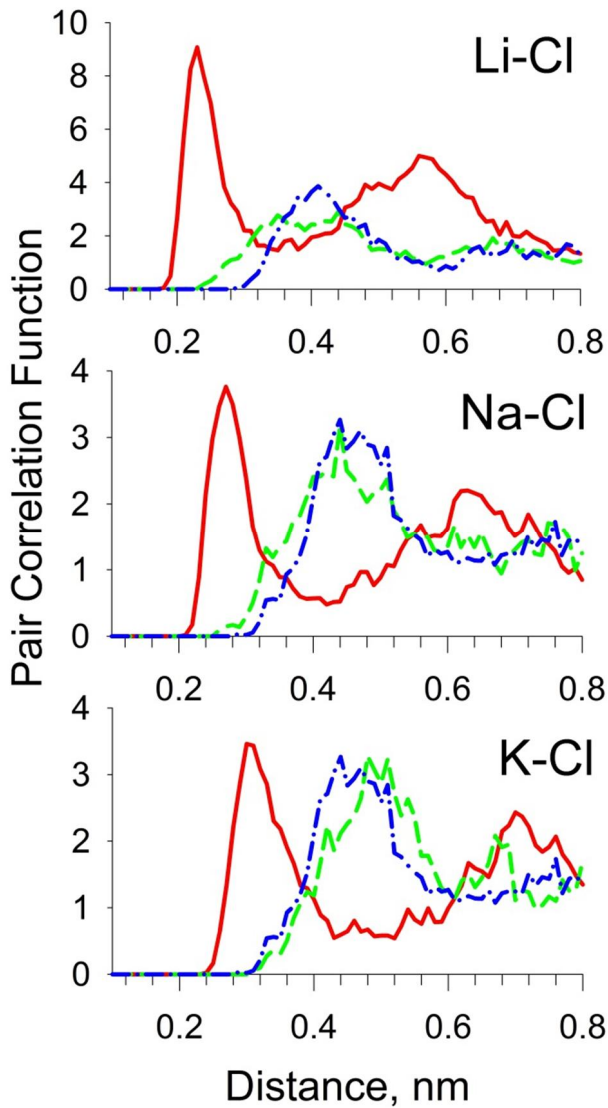

Figure 6. Pair correlation functions in the simulated ionic clusters derived at 1500 K: cation-anion (red solid line), cation-cation (green dashed line), and anion-anion (blue dash-dotted line).

Unlike a ball, an atom does not possess a fixed radius. The only way to assign certain number is by measuring the distance between two identical nuclei and then dividing the result by two. The tabulated ionic radii are 0.076 nm for $Li^+$, 0.102 nm for $Na^+$, 0.138 nm for $K^+$, and 0.181 nm for $Cl^-$. The tabulated covalent radii are 0.128 nm for Li, 0.166 nm for Na, 0.203 nm for K, and 0.102 nm for Cl. That is, the prevailing ionic bonding in the clusters would result in the PCF first peak positions at 0.257 (LiCl), 0.283 (NaCl), and 0.319 nm (KCl). In turn, the prevailing covalent bonding in the clusters would result in the PCF first peak positions at 0.230 (LiCl), 0.268 (NaCl), and 0.305 nm (KCl). Compare these values with the actual positions of the first peaks at 300 K: 0.23 (LiCl), 0.27 (NaCl), and 0.31 nm (KCl). Interestingly, the positions the first peaks at 1500 K are same, although the heights of these peaks are significantly smaller. Thus, temperature induced changes in the bonding nature are marginal.

Figure 7 compares distribution of point ESP charges in the optimized geometries of $Li_{10}Cl_{10}$, $Na_{10}Cl_{10}$, and $K_{10}Cl_{10}$. An entropic contribution was deliberately excluded to observe symmetry in these structures. An ideal symmetry would imply equal fractional charges on all cations and all anions, as it is the case in salt crystals (periodic systems). The symmetry is not ideal in small ionic clusters, as Figure 7 evidences. Interestingly, the ESP charges are most uniform in $Li_{10}Cl_{10}$, which is in line with a stronger binding energy in this system.

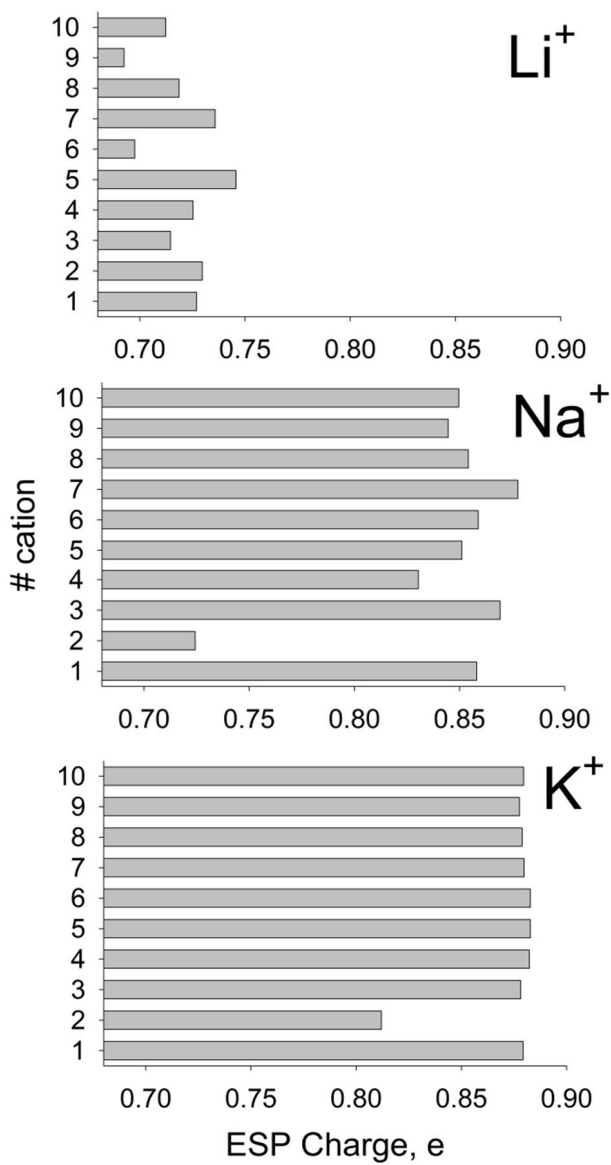

Figure 7. Distribution of electrostatic potential fitted charges over cations in the optimized ionic clusters of $Li_{10}Cl_{10}$, $Na_{10}Cl_{10}$, and $K_{10}Cl_{10}$. The y-axis indicates a serial number of the cation irrespective of its spatial localization and bonding in the ionic cluster.

**Conclusions**

Using Born-Oppenheimer molecular dynamics simulations based on pure density functional theory, we investigated interaction non-additivities at room and elevated temperatures. The interaction non-additivities are significant in all considered systems at all conditions making fractional electrostatic charges deviate from unity. We showed that temperature exhibits a minor impact on electron delocalization in the small inorganic ionic clusters, $Li_{10}Cl_{10}$, $Na_{10}Cl_{10}$,

$K_{10}Cl_{10}$. A stronger trend was found in respect to cation nature. Dipole moments increase systematically upon heating due to growing thermal fluctuations of the corresponding structures and thermal expansion. Fractional point charges remain largely unaltered, in turn.

The provided analysis is based on the small clusters simulated in vacuum rather than condensed phase. This is due to a set of methodological considerations. First, computational cost for simulations of quantum dynamics is often prohibitive for systems of more than a few hundred electrons. Second, computation of the long-range electrostatic potential is not straightforward in periodic systems. In addition, point charges cannot be univocally defined far from the surface of an ion cluster/aggregate (the "buried atom" problem). Third, the dipole moment is most meaningful in the case of a single ion pair, whereas higher-order electric moments suit better for description of large molecular formations. We expect that periodic systems would exhibit somewhat different behavior due to elimination of surface effects. Nevertheless, the qualitative conclusions of this work should not change.

**Acknowledgments**

This research has been partially supported by grant CHE-1300118 from the US National Science Foundation. V.V.C. is CAPES professor in Brazil under the "Science Without Borders" initiative.

**Author Information**

E-mail address for correspondence: vvchaban@gmail.com (V.V.C.)

TOC Graphic

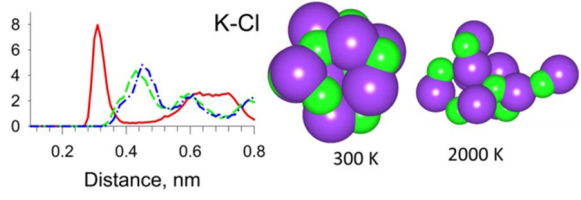